\documentclass[aps,twocolumn,prd,showpacs,nofootinbib,floatfix]{revtex4-1}
\usepackage{latexsym}
\usepackage{psfrag}
\usepackage{graphicx,amsmath,amsfonts,amssymb,dcolumn,epsfig,bm,float}
\usepackage{epsfig}%
\usepackage{color}
\usepackage{graphics}
\usepackage{epstopdf}
\usepackage[margin=0.5in]{geometry}
\usepackage[hang,small,bf]{caption}
\usepackage{rotating}
\usepackage{subfigure}
\usepackage{silence}
\usepackage{appendix}
\usepackage{tabularx}
\usepackage{multirow}
\usepackage{pifont}
\usepackage{lipsum}
\def\beq{\begin{equation}}
\def\eeq{\end{equation}}
\def\bea{\begin{eqnarray}}
\def\eea{\end{eqnarray}}

\begin{document}

\title{Quantum discord as a tool for comparing collapse models and decoherence}

\author{Shreya Banerjee} \email{shreya.banerjee@tifr.res.in}

\author{Sayantani Bera} \email{corresponding author: sayantani.bera@tifr.res.in}

\author{Tejinder  P. Singh} \email{tpsingh@tifr.res.in}
\affiliation{ Tata Institute of Fundamental Research \\
Homi Bhabha Road, Mumbai 400005, India}
\date{\today}
\begin{abstract}
  \noindent 
The quantum to classical transition maybe caused by decoherence or by dynamical collapse of the wave-function. We propose quantum discord as a tool, 1) for comparing and contrasting the role of a collapse model (Continuous Spontaneous Localization) and various sources of decoherence (environmental and fundamental), 2) for detecting collapse model and fundamental decoherence for an experimentally demonstrated macroscopic entanglement. We discuss the experimental times which will lead to the detection of either Continuous Spontaneous Localization or fundamental decoherence. We further put bounds on the collapse parameters from this experiment for quantum discord.


\end{abstract}

\maketitle


\section{Introduction}
The nature of the quantum to classical transition is not yet fully understood. The apparent absence of macroscopic position superpositions is not explained by the Schr\"odinger equation, and some additional input is needed. This loss of superposition may be caused by environmental decoherence (if supplemented by the many worlds interpretation). Alternatively the explanation may lie in Bohmian mechanics, or in fundamental decoherence induced by gravity, or it may result from collapse models which modify the Schr\"odinger equation \cite{revbassi}.\\
 If one creates a macroscopic superposition, it is useful to have a measure of its `quantumness' or lack thereof. Various such measures have been proposed in the literature (for a review see for instance \cite{jeong}). Another significant measure of quantumness is quantum correlation, an important instance of which is entanglement, whereby superposed states of one quantum system are entangled with the superposed states of another. Various measures of
quantum correlation, which quantify the extent of departure from classicality, have also been proposed \cite{vedrev}.
One such measure, quantum discord \cite{zurek, vedral}, is the subject of the present paper.  We use discord as a means of
comparing collapse model and gravity induced decoherence in a specific system for which experimentalists succeeded in creating a macroscopic entangled state \cite{science} and minimising environmental effects. The extent and evolution of discord helps to discriminate between them.\\

 We consider an experiment in which an entanglement was created between two diamond pieces separated by a macroscopic distance of about 15 cm. Each diamond is set up in a superposition of two quantum states, which entangle with the states of the other diamond, with the entanglement lasting about $10^{-12}$ sec.  In a sense, this is an example of macroscopic quantum behaviour, and it is important to ask how this observation constrains the parameters of collapse models such as Continuous Spontaneous Localization (CSL) \cite{Pearle:89, Ghirardi2:90} which predict breakdown of superposition in macroscopic systems. This study has been recently carried out in \cite{diamondbassi} and it has been found that the CSL parameters are only weakly constrained by this experiment, with stronger bounds coming from matter wave interferometry, spontaneous X-ray emission, and millikelvin-cooled nanocantilevers (see \cite{Vinante2016} and especially Fig. 3 therein). The two phenomenological parameters of CSL are the rate constant $\lambda$ which determines the rate of spontaneous collapse, and the physical length scale $r_C$ to which the collapsing wave-function gets confined. A typical choice for $r_C$ is $10^{-5}$ cm, while the rate constant in the original CSL model was suggested to be $\lambda_{CSL}\sim 10^{-17}$ s$^{-1}$. A higher value of $10^{-8}$ s$^{-1}$ for the rate constant has been suggested by Adler \cite{Adler}. The authors in \cite{diamondbassi} find that the entanglement observed in the diamond system puts an upper bound of about $10^{-3}$ s$^{-1}$ on the rate constant, which is significantly higher than the theoretically proposed value.\\ 

 In the present paper we use discord as a means to ask: to what extent the duration of the entanglement should be enhanced, so that the experiment in question can reach and probe an interesting domain of values for the CSL rate constant, not reached yet by other experiments? We ask a similar question also for fundamental gravity induced decoherence. As done in \cite{diamondbassi}, here also we try to put bounds on CSL parameters using quantum discord and show that these two studies are consistent.\\

 We plan to follow up this work with application of quantum discord in other studies (where environmental decoherence also plays an important role) related to the quantum measurement problem. Of particular interest is to employ a discord analysis in applications of CSL to the inflationary universe \cite{inflation}, thereby generalizing a recent novel application of discord to the anisotropies of the cosmic microwave background \cite{Martin}.\\
 
 Section \ref{secdiscord} of this paper briefly recalls the definition of quantum discord. Section \ref{experimentalsetup} recalls the experimental setup of \cite{science}, while sections \ref{csl}-\ref{envd} compute the discord for CSL, gravitational decoherence and environmental decoherence respectively. Section \ref{results} compares discord for collapse and gravity induced decoherence, and discusses the feasibility of testing these models by carrying out the said experiment for a longer time.
We have also shown the allowed $\lambda_{CSL}-r_{C}$ parameter space as obtained from this experiment.

\label{intro}

\section{Quantum Discord : a brief overview}
\label{secdiscord}
In the context of quantum information theory, quantum discord is a measure of the quantumness or non-classicality between two bipartite subsystems of a quantum system. It was first introduced by the authors in \cite{zurek} and \cite{vedral}, where they showed that the vanishing of discord can be considered as a criterion for determining the pointer states of a system. 

Mathematically, quantum discord is defined in terms of the difference between two measures of mutual information between two subsystems, which coincide for classical correlations. As we know from quantum information theory, the quantum correlation/mutual information between two subsystems $X$ and $Y$ is given by
\begin{equation}
\mathcal{I}(X : Y) = S(\hat{\rho}_X) + S(\hat{\rho}_Y) - S(\hat{\rho})
\end{equation}
where $S(\hat{\rho}_X)$ and $S(\hat{\rho}_Y)$ are the von-Neumann entropies for the subsystems $X$ and $Y$ respectively. $S(\hat{\rho})$ is the joint quantum entropy for the whole system. The general definition for von-Neumann entropy for any system with density matrix $\hat{\rho}$ is given by
\begin{equation}
S(\hat{\rho}) = -{\rm Tr}\hat{\rho} \ln\hat{\rho} = -\sum_i\sigma_i \ln \sigma_i
\end{equation}
Here $\sigma_i$ are the eigenvalues of $\hat{\rho}$.

An alternative definition of mutual information between two subsystems is obtained as a result of the information gained about one subsystem due to the measurement on the other subsystem. Let us imagine measuring an observable of the subsystem $Y$ through the projection operator $\hat{\Pi}_j$. Under such a scenario, the mutual information between the two subsystems can be mathematically expressed as,
\begin{equation}
\mathcal{J}(X : Y) = S(\hat{\rho}_X) - \sum_j p_j S(\hat{\rho}_{(X;\hat{\Pi}_j)})
\end{equation}
where $\sum_j p_j S(\hat{\rho}_{(X;\hat{\Pi}_j)})$ is the conditional quantum entropy of $X$ for a  given state of $Y$. Here, 
\begin{eqnarray}
\label{conditional}
\hat{\rho}_{(X;\hat{\Pi}_j)} &&=  {\rm Tr}_Y\left(\frac{\hat{\Pi}_j \hat{\rho}\hat{\Pi}_j }{p_j}\right)\\
p_j &&= {\rm Tr}(\hat{\Pi}_j \hat{\rho}\hat{\Pi}_j) \quad\quad({\rm probability\ of\ getting\ a\ state\ of} \nonumber \\ &&~~~~~~~~~~~~~~~~~~~~~~~~~~~   X\ {\rm given}\ j^{th}\ {\rm state\ of}\ Y)
\end{eqnarray}
In Eq. \eqref{conditional}, $\hat{\Pi}_j = |j\rangle \langle j|$ is the projection operator which is used to perform a measurement on the subsystem $Y$ and $j$ represents the different outcomes of this measurement.

Classically, using Bayes' theorem, these two measures of information are identical. However, for quantum systems, this need not be true. The deviation from this equality gives the measure of quantum discord as defined below:
\begin{equation}
\delta(X : Y) = \mathcal{I}(X : Y) - \mathcal{J}(X : Y)
\label{quantumdiscord}
\end{equation}
$\delta = 0$ corresponds to classical subsystems and $\delta > 0$ shows the presence of quantum correlation between two subsystems. In general, discord is asymmetric under the interchange of the systems $X$ and $Y$, but in our present work the two subsystems are identical.

\section{Experimental and Theoretical Set-Up}
\label{exptheo}
\subsection{Basic Experimental set-up}
\label{experimentalsetup}
Our analysis is based on the experiment reported in \cite{science} where the authors created an entanglement between two macroscopic diamond pieces separated by a large distance ($15$ cm) at room temperature and verified that this entanglement is preserved during the experimental time ($\sim 10^{-12} s$). The idea is to excite phonon modes within the two diamond crystals using Raman scattering such that the phonon modes of the two spatially separated crystals become entangled. These phonon modes are actually macroscopic and thus preserving the quantum entanglement between these modes ensures the presence of entanglement in the macro regime. The experimental set-up consists of a source generating ultra-short optical pulses which are used to excite or de-excite phonon modes within the crystal via the emission of Stokes or anti-Stokes photons respectively. The pulse (also called the pump pulse) is split into two parts using a $50:50$ beam-splitter and sent to the two crystals (denoted by ``L" and ``R") simultaneously to create an entanglement between the two crystals. This entire process generates two Stokes modes at the two crystals which are then made to interfere with a phase difference $\phi_s$ created using a half-wave plate and a polarizer. The interfered Stokes modes are detected at the detector. Thus, detection of Stokes photons at the detector indicates the fact that the two crystals are now entangled. To verify that this entanglement is preserved until environmental decoherence effects destroy the entanglement, a second pulse, known as the probe pulse, is sent to each of the crystals simultaneously again using the $50:50$ beam splitter. Due to the interaction with this probe pulse, some of the excited phonon modes within the crystals get de-excited to the ground state resulting in an emission of anti-Stokes photons. Finally, these anti-Stokes photons, coming out from the L and R crystals, are also made to interfere in a similar manner with a phase difference $\phi_a$ and detected at another detector. Thus, by detecting Stokes and anti-Stokes photons at the two detectors, one concludes that a macroscopic entanglement is created and preserved within the time scale of the experiment. For a more detailed description of the experiment, one may refer to \cite{science}. 

The excited phonon modes induce displacements of the atoms within the crystals such that two neighbouring atoms oscillate in the opposite directions (``out of phase" oscillation), thus creating two counter-oscillating sublattices within each crystal. We denote $|0 \rangle$ as the state when no phonon modes are excited. Thus, $|0_L \rangle$ and $|0_R \rangle$ represent the ground states of L and R crystals respectively and $|1_L \rangle$ and $|1_R \rangle$ represent the excited phonon states of L and R crystals respectively. The basis for the entangled system is thus: $|0_L, 0_R \rangle$, $|1_L, 0_R \rangle$, $|0_L, 1_R \rangle$, $|1_L, 1_R \rangle$. Further theoretical calculations will be done using this basis. 
\subsection{Discord in the presence of collapse due to CSL}
\label{csl}

In this paper we study quantum discord for the above experimental setup, in the presence of a mechanism responsible for spontaneous dynamical collapse of the superposed state. We apply by far the most realistic and advanced collapse model for multiparticle indistinguishable systems, known as Continuous Spontaneous Localization or CSL. Originally proposed by \cite{Pearle:89,Ghirardi2:90},  CSL modifies quantum mechanics to include an anti-hermitian, norm-preserving,  stochastic and non-linear component in the Schr\"{o}dinger equation. Thus it possesses the important properties of localization, amplification and dynamically induced collapse. As has been already mentioned in \cite{diamondbassi}, in the present case, CSL tries to prevent the superposition of two different matter distributions (diamond lattices at rest or counter oscillating) in space.  For a system of particles, the mass-proportional version of CSL equation in position basis in Fock space is given by
\begin{equation} 
\begin{split}
\label{eq:csl-massa}
& d\psi_t = \Big[-\frac{i}{\hbar}\hat{H}dt 
 + \frac{\sqrt{\gamma}}{m_{0}}\int d\mathbf{x} (\hat{M}(\mathbf{x}) - \langle \hat{M}(\mathbf{x}) \rangle_t)
dW_{t}(\mathbf{x})  \\ &
 ~~~~~~~~~~ - \frac{\gamma}{2m_{0}^{2}} \int d\mathbf{x}\, d\mathbf{y}\, \mathcal{G}({\bf x}-{\bf y})
(\hat{M}(\mathbf{x})-\langle \hat{M}(\mathbf{x}) \rangle_t) \times \\ & ~~~~~~~~~~~~~~~~~~~~~~~~~~~~(\hat{M}(\mathbf{y})-\langle \hat{M}(\mathbf{y}) \rangle_t)dt\Big] \psi_t  
\end{split}
\end{equation}
The linear part is governed by $H$ - the standard quantum Hamiltonian of the system, and the other two terms induce the collapse of the wave function. The mass $m_0$ is a reference mass, which is taken equal to that of a nucleon. Here $\mathcal{G}({\bf x}-{\bf y})$ is the spatial two point correlation function for the Gaussian noise. The noise two point correlation for the Wiener process is given by $\langle W_{t_1}(\mathbf{x})W_{t_2}(\mathbf{y})\rangle = \mathcal{G}({\bf x}-{\bf y}) \delta (t_1 -t_2)$. Thus the noise two point function is Dirac delta in time and Gaussian in space \cite{revbassi}. $\gamma$ is the positive coupling constant which sets the strength of the collapse process, while $\hat{M}({\bf x})$ is a smeared mass density operator given by
\begin{eqnarray}
\hat{M}({\bf x})&=& \sum_{j}m_{j}\hat{N}_{j}({\bf x}) \\
\hat{N}_{j}({\bf x})&=& \int d{\bf y}g({\bf y}-{\bf x})\hat{a}_{j}^{\dagger}({\bf y})\hat{a}_{j}({\bf y})
\end{eqnarray}
Here $\hat{a}_{j}^{\dagger}({\bf y})$ and $\hat{a}_{j}({\bf y})$ are, respectively, the creation and annihilation operators of a particle of type $j$ at the space point ${\bf y}$. The smearing function is defined as
\begin{equation}
g({\bf x})=e^{-{\bf x}^{2}/2r_{C}^{2}}
\end{equation} 
where $r_{C}$ is another new phenomenological constant.

$W_t({\bf x})$ is an ensemble of independent Wiener processes.  The wave-function is now a stochastic entity, and one can construct a Lindblad type linear master equation for the corresponding density matrix. The CSL master equation in position space is given by \cite{bassicsl}
 \begin{equation}
 \frac{d\hat{\rho}(t)}{dt}=-\frac{i}{\hbar}[\hat{H},\hat{\rho}(t)]-\frac{\lambda_{\rm CSL}}{2r_{C}^{3}\pi^{3/2}m_{0}^{2}}\int d{\bf x}[\hat{M}({\bf x}),[\hat{M}({\bf x}),\hat{\rho}(t)]]
 \label{master}
 \end{equation}
 where $\lambda_{\rm CSL}=\gamma/(4\pi r_{C}^{2})^{3/2}$. Here $\hat{\rho}(t)$ is the total density matrix of the entire system.

For the experimental setup under consideration, we analyse the effect of CSL on each diamond crystal L and R. Each crystal can either be at rest or oscillating (neighbouring atoms oscillating in opposite directions). The effect of CSL is to make each crystal collapse into either of these states. This affects the entanglement and the quantum discord between the two crystals.

Calculations in this section are based on the analysis done in \cite{diamondbassi}, where the authors have put bounds on CSL collapse parameters by studying the entanglement between the two diamond crystals in the presence of CSL. Similarly, here, we try to examine the effect of CSL on the quantum discord between the two crystals.

 Rewriting Eq. \eqref{master} for the experimental setup described in section  \ref{experimentalsetup}, we get
 \begin{equation}
\label{master2}
 \frac{d}{dt}  \hat{\rho}(t)  = - \frac{i}{\hbar}\,  [ \hat{H},\hat{\rho}(t) ] - 2 \eta [\hat {q}^L, [\hat{q}^L, \hat{\rho}_t]]  -  2 \eta [\hat {q}^R, [\hat{q}^R, \hat{\rho}(t) ]] 
\end{equation}
where $\hat {q}^L$ and $\hat {q}^R$ are the position operators for the left and right diamond systems respectively.  Here $\eta$ is a function of the mass distribution for the two sublattices (the mass distribution is same for the two sublattices).

In the above experiment, let us consider each diamond subsystem to be of cylindrical shape with radius $R$ and width $d$. Let $N$ be the total number of atoms in the sublattice contributing to the phonon modes. The mass $m$ of each sublattice is given by $m= 12\times N \times m_0$ ($12$ is the atomic number of the carbon atom). It has been shown in \cite{diamondbassi}, taking into consideration the above parameters for the experimental system, the form of $\eta$ and hence the mass distribution for each sublattice is given by the general expression
\begin{equation}
\eta = \lambda_{\rm CSL} \frac{N^2}{d^2} \Gamma_{\perp}\left(\frac{R}{\sqrt{2}\, r_C}\right)\, \left[ 1 - e^{- \frac{d^2}{4\, r_C^2}} \right]
\label{eta}
\end{equation}
where:
\begin{equation} 
\Gamma_{\perp} (x)= \frac{2}{x^2} \left[ 1- e^{-x^2} \, \left( I_0(x^2) +I_1(x^2) \right) \right],
\end{equation}
with $I_0$ and $I_1$ being the modified Bessel functions. For a detailed derivation of the above form of $\eta$, one may refer to \cite{diamondbassi}.

In order to solve Eq. \eqref{master2}, we express the position operators in terms of creation and annihilation operators, for both the left and right sublattices:
\begin{equation}
\hat {q}^L  =\sqrt{\frac{\hbar}{6\omega m _0}} \frac{\hat a_L + \hat a_L^{\dagger}}{\sqrt{2}}\,, \qquad \hat {q}^R =\sqrt{\frac{\hbar}{6\omega m_0}} \frac{\hat a_R + \hat a_R^{\dagger}}{\sqrt{2}}  \,.
\end{equation}
where $\omega$ is the frequency of the phonon.

Using the basis states as has been described in section \ref{experimentalsetup} and following the analysis of \cite{diamondbassi}, we can compute the matrix elements of the density matrix using the following equation:

\begin{widetext}
\begin{small}
\begin{align}
\left( \begin{matrix} 
 \dot{\rho}_{11}\ & \dot{\rho}_{12}\ & \dot{\rho}_{13}\ & \dot{\rho}_{14} \\
\dot{\rho}_{21}\ & \dot{\rho}_{22}\ & \dot{\rho}_{23}\ & \dot{\rho}_{24} \\
\dot{\rho}_{31}\ & \dot{\rho}_{32}\ & \dot{\rho}_{33}\ & \dot{\rho}_{34} \\
\dot{\rho}_{41}\ & \dot{\rho}_{42}\ & \dot{\rho}_{43}\ & \dot{\rho}_{44} \\
 \end{matrix}\right)= -i\omega\left( \begin{matrix} 
 0\ & -\rho_{12}\ & -\rho_{13}\ & -2 \, \rho_{14} \\
\rho_{21}\ & 0\ & 0\ & - \rho_{24} \\
\rho_{31}\ & 0\ & 0\ & - \rho_{34} \\
2 \, \rho_{41}\ & \rho_{42}\ & \rho_{43}\ & 0 \\
 \end{matrix}\right)-\frac{1}{2}\Lambda \left( \begin{matrix} 2 \rho_{11}-\rho_{22}-\rho_{33}\ \ & 2 \rho_{12}-\rho_{21}-\rho_{34}\ \ & 2 \rho_{13}-\rho_{24}-\rho_{31}\ \ & 2 \rho_{14}-\rho_{23}-\rho_{32} \\ 2 \rho_{21}-\rho_{12}-\rho_{43}\ \ & 2 \rho_{22}-\rho_{11}-\rho_{44}\ \ & 2 \rho_{23}-\rho_{41}-\rho_{14}\ \ & 2 \rho_{24}-\rho_{13}-\rho_{42}  \\ 2 \rho_{31}-\rho_{42}-\rho_{13}\ \ & 2 \rho_{32}-\rho_{41}-\rho_{14}\ \ & 2 \rho_{33}-\rho_{44}-\rho_{11}\ \ & 2 \rho_{34}-\rho_{43}-\rho_{12} \\ 2 \rho_{41}-\rho_{32}-\rho_{23}\ \ & 2\rho_{42}-\rho_{31}-\rho_{24}\ \ & 2\rho_{43}-\rho_{34}-\rho_{21}\ \ & 2 \rho_{44}-\rho_{33}-\rho_{22} \end{matrix}\right) \nonumber \\
 \label{densitymatrix}
 \end{align}
 \end{small}
\end{widetext} 
where 
\begin{equation}
\Lambda=\frac{2\eta\hbar}{3\omega m_0}
\label{lambda}
\end{equation}
Here index $1$ represents $|0_{L},0_{R}\rangle$, $2$ represents $|1_{L},0_{R}\rangle$, $3$ stands for $|0_{L},1_{R}\rangle$ and $4$ represents $|1_{L},1_{R}\rangle$.

In Eq. \eqref{densitymatrix}, the first matrix represents the time derivative of the elements of the density matrix. The second matrix corresponds to the elements of the term $[\hat{H},\hat{\rho}(t)]$ and the third matrix represents the elements of $[\hat {q}^L,[\hat{q}^L, \hat{\rho}_t]]+[\hat{q}^R,[\hat{q}^R, \hat{\rho}(t)]]$. 

Using 
\begin{equation}
\Psi_{0}(t)=\frac{1}{\sqrt{2}}|1_{\rm L},0_{\rm R}\rangle+\frac{1}{\sqrt{2}}|0_{\rm L},1_{\rm R}\rangle
\end{equation}
as the initial wavefunction of the entire system, we solve the matrix equation given by Eq. \eqref{densitymatrix} for each element of $\rho$. The complete expression for the solution of $\rho$ in matrix form is given by
\begin{eqnarray}
\rho=\left( \begin{matrix} 
 \rho_{11}\ & 0 \ & 0 \ & \rho_{14} \\
0\ & \rho_{22}\ & \rho_{23} \ & 0 \\
0\ & \rho_{32}\ & \rho_{33}\ & 0 \\
\rho_{41}\ & 0\ & 0\ & \rho_{44} \\
 \end{matrix}\right)
 \end{eqnarray}
 where
\begin{widetext}
\begin{align}
\rho_{11}=\rho_{44}= \frac{1}{4}\left(1-e^{-2\Lambda t}\right);~~~~~~~~~~~~~~~~~~~~~~~~~~~~~~~~~~~~~~~~~~~~~~~~~~~~~~~~~~~~~~~~~~~~~~~~~ \\
\rho_{14}=\frac{\Lambda}{4(\Lambda^{2}-4 \omega^{2})^{3/2}}e^{[-\Lambda t-\alpha t]}\left(-e^{\Lambda t}\Lambda^{2}+\Lambda^{2}e^{[\Lambda t+\beta t+\alpha t]}+4 \omega^{2}e^{\Lambda t}-4 \omega^{2}e^{[\Lambda t+\beta t+\alpha t]}\right.\nonumber \\  \left. +2i\omega\sqrt{\Lambda^{2}-4\omega^2} e^{\Lambda t}-4i\omega\sqrt{\Lambda^{2}-4\omega^2} e^{\alpha t}+2i\omega\sqrt{\Lambda^{2}-4\omega^2} e^{[\Lambda t+\beta t+\alpha t]}\right); \\
\rho_{22}=\rho_{33}=\frac{1}{4}\left(1+e^{-2\Lambda t}\right);~~~~~~~~~~~~~~~~~~~~~~~~~~~~~~~~~~~~~~~~~~~~~~~~~~~~~~~~~~~~~~~~~~~~~~~~~ \\
\rho_{23}=\rho_{32}=\frac{e^{[-\Lambda t-\alpha t]}\left(\Lambda^{2}e^{\Lambda t}+\Lambda^{2}e^{[\Lambda t+\beta t+\alpha t]}-8\omega^{2}e^{\alpha t}\right)}{4(\Lambda^{2}-4\omega^{2})};~~~~~~~~~~~~~~~~~~~~~~~~~~~~~~~~~~~ \\
\rho_{41}=\frac{\Lambda}{4(\Lambda^{2}-4 \omega^{2})^{3/2}}e^{[-\Lambda t-\alpha t]}\left(-e^{\Lambda t}\Lambda^{2}+\Lambda^{2}e^{[\Lambda t+\beta t+\alpha t]}+4 \omega^{2}e^{\Lambda t}-4 \omega^{2}e^{[\Lambda t+\beta t+\alpha t]}\right.\nonumber \\  \left. -2i\omega\sqrt{\Lambda^{2}-4\omega^2} e^{\Lambda t}+4i\omega\sqrt{\Lambda^{2}-4\omega^2} e^{\alpha t}-2i\omega\sqrt{\Lambda^{2}-4\omega^2} e^{[\Lambda t+\beta t+\alpha t]}\right) 
\end{align}

\end{widetext}
Here $\alpha=\Lambda+\sqrt{\Lambda^{2}-4\omega^{2}}$ and $\beta=-\Lambda+\sqrt{\Lambda^{2}-4\omega^{2}}$.

By diagonalizing the above density matrix, one gets the four eigenvalues for $\rho$ as
\begin{eqnarray}
\sigma_{1}&=&\rho_{22}-\rho_{23};\ \sigma_{2}=\rho_{22}+\rho_{23};\nonumber \\ \sigma_{3}&=&\rho_{11}-\sqrt{\rho_{14}\rho_{41}};\ \sigma_{4}=\rho_{11}+\sqrt{\rho_{14}\rho_{41}} 
\end{eqnarray} 
In order to find the density matrices for the sublattices/subsystems (known as reduced density matrix), one needs to trace out the degrees of freedom of either of the sublattices. Thus, by tracing over the left (right) sublattice, we get the reduced density matrix for the right (left) sublattice. The expressions for the reduced density matrices are given by
\begin{eqnarray}
\rho_{L}=\rho_{R}=\frac{1}{2}\left( \begin{matrix} 
 1\ \ \ & 0 \\
0\ \ \ &  1 \\
 \end{matrix}\right)
 \end{eqnarray}
 
 Using the definition for von Neumann entropy as already given in section \ref{secdiscord}, we get the expressions for the entropies of the different subsystems and the total system as
 \begin{eqnarray}
 S(\hat{\rho})= -\sum_{i}\sigma_{i}\ln \sigma_{i};\ S(\hat{\rho}_{L})=S(\hat{\rho}_{R})=-\ln \left(\frac{1}{2}\right)
 \end{eqnarray}
 Using the definition of the projection operator $\hat{\Pi}_{j}$ given in section \ref{secdiscord}, we get the expression for the quantity $\sum_{j}p_{j}S(\hat{\rho}_{(L;\hat{\Pi}_j)})$ as
 \begin{eqnarray}
\sum_{j}p_{j}S(\hat{\rho}_{(L;\hat{\Pi}_j)})&=&-\frac{1}{4}\left(1+e^{-2\Lambda t}\right)\ln \left\vert \frac{(1+e^{-2\Lambda t})}{2}\right\vert \nonumber \\ &&~~~ -\frac{1}{4}\left(1-e^{-2\Lambda t}\right) \ln \left\vert \frac{(1-e^{-2\Lambda t})}{2}\right\vert \nonumber \\
\end{eqnarray} 
Therefore, using Eq. \eqref{quantumdiscord}, the expression for quantum discord as a function of time is given by
\begin{equation}
\begin{split}
& \delta(L:R)=S(\hat{\rho}_{R})-S(\hat{\rho})+\sum_{j}p_{j}S(\hat{\rho}_{(L;\hat{\Pi}_j)})  \\
&~~~~~= -\ln \left(\frac{1}{2}\right)+\sum_{i}\sigma_{i}\ln \sigma_{i}-\frac{1}{4}\left(1+e^{-2\Lambda t}\right)\ln \left\vert \frac{(1+e^{-2\Lambda t})}{2}\right\vert \\ &~~~ -\frac{1}{4}\left(1-e^{-2\Lambda t}\right) \ln \left\vert \frac{(1-e^{-2\Lambda t})}{2}\right\vert 
\label{discord}
\end{split}
\end{equation} 
In Section \ref{results}, we will compute this quantity and graphically compare its time evolution with other models.

\subsection{Discord in the presence of gravity induced decoherence}
\label{gid}

 Whether gravity can play a role in the decoherence mechanism of an object was first studied by Karolyhazy and this was later on taken up by others \cite{k1,k2}. A similar (but slightly different) approach was followed by  Di\'osi and Penrose \cite{diosi1,diosi2,penrose} where they suggested that the difference in the self-gravitational energy of an object in the two possible mass configurations leads to decoherence between the possible states. The importance of this gravity induced decoherence, if it is present, is that it is intrinsic to the system and cannot be eliminated unlike environmental decoherence. Thus it is important to study this effect along with the collapse models.

Here we consider the model proposed by  Di\'osi (also known as Di\'osi - Penrose model or DP model). The dynamics of the system with a mass $m$ is governed by the master equation of the following form:
\begin{equation}
\begin{split}
\frac{\partial\rho(x,y,t)}{\partial t}= & \frac{i\hbar}{2m}(\nabla_x^2-\nabla_y^2)\rho(x,y,t)\\ 
& ~~~~~-\frac{1}{\hbar}\left[U(x-y)-U(0)\right]\rho(x,y,t)
\end{split}
\label{diosimastereqn}
\end{equation}
The first term on the right hand side is the usual quantum mechanical evolution and the second term represents the decay of the off-diagonal terms due to the  decoherence mechanism. Here, $U(x,y)$ denotes the Newtonian potential energy between the two mass configurations at $x$ and $y$. If one considers the mass distribution as a homogeneous sphere of radius $R'$ and mass $m$ then this potential takes the form,
\begin{eqnarray}
U(x,y)&&= -\frac{G m^2}{R'}\left(\frac{6}{5}-\frac{1}{2}\frac{|x-y|^2}{R'^2}\right) \quad\quad\quad  |x-y| \ll R'  \nonumber \\
&&= -\frac{G m^2}{|x-y|} \quad\quad\quad\quad\quad\quad\quad\quad\quad \quad\quad    |x-y| \gg R' \nonumber \\
\end{eqnarray}

For the present setup, $U(x,y)$ denotes the Newtonian potential between the rest and the oscillating mass configurations of each diamond crystal. Thus, like CSL, here, we are examining the effect of gravity induced decoherence on each crystal. Since the displacement of the sublattices are very small compared to the size of the crystals, we take the first form above, for $|x-y| \ll R'$. It is useful for our purpose, to write the above master equation in operator form as given below. In position basis, we have,

\begin{eqnarray}
\frac{\partial\langle x|\hat{\rho}|y \rangle}{\partial t}&&= -\frac{i}{\hbar}\langle x|[\hat{H},\hat{\rho}]|y\rangle - \frac{G m^2}{2\hbar R'^3}(x-y)^2\langle x|\hat{\rho}|y \rangle \nonumber \\
&&= -\frac{i}{\hbar}\langle x|[\hat{H},\hat{\rho}]|y\rangle \nonumber \\ && - \frac{G m^2}{2\hbar R'^3}\left(\langle x|\hat{q}\hat{q}\hat{\rho}|y \rangle - 2\langle x|\hat{q}\hat{\rho}\hat{q}|y \rangle + \langle x|\hat{\rho}\hat{q}\hat{q}|y \rangle\right)\nonumber \\
\label{masterop}
\end{eqnarray}
where we have used the fact that $\hat{q}|x \rangle = x|x \rangle$ and $\hat{q}|y \rangle=y|y \rangle$.
Eq. $\eqref{masterop}$ can be written as,
\begin{equation}
\frac{\partial\langle x|\hat{\rho}|y \rangle}{\partial t}= -\frac{i}{\hbar}\langle x|[\hat{H},\hat{\rho}]|y\rangle - \frac{G m^2}{2\hbar R'^3}\langle x|[\hat{q},[\hat{q},\hat{\rho}]]|y \rangle
\end{equation}
Thus, in operator form, we have
\begin{equation}
\frac{\partial\hat{\rho}}{\partial t}= -\frac{i}{\hbar}[\hat{H},\hat{\rho}] - \frac{G m^2}{2\hbar R'^3}[\hat{q},[\hat{q},\hat{\rho}]]
\label{diosimasterop}
\end{equation}
This form of the master equation has the same structure as in CSL model discussed earlier. For the system under study having frequency $\omega$ and consisting of two crystals L and R, we can write 
\begin{equation}
\frac{\partial\hat{\rho}}{\partial t}= -\frac{i}{\hbar}[\hat{H},\hat{\rho}] - \frac{G m^2}{2\hbar R'^3}[\hat{q}^L,[\hat{q}^L,\hat{\rho}]]- \frac{G m^2}{2\hbar R'^3}[\hat{q}^R,[\hat{q}^R,\hat{\rho}]]
\label{diosimasterop1}
\end{equation} 
Comparing this form with the master equation given in Eq. $\eqref{master2}$, we have,
\begin{equation}
\eta_{Diosi}=\frac{G m^2}{4\hbar R'^3}
\end{equation}
The decay rate $\Lambda_{Diosi}$ for this model is thus given by,
\begin{equation}
\Lambda_{Diosi}=\frac{Gm^2}{6\omega R'^3 m_0}
\label{lamdiosi}
\end{equation}

The expression for discord for this model will again be given by Eq. \eqref{discord} with $\Lambda$ substituted by $\Lambda_{Diosi}$. Using this final form we will compare its time evolution with other models in Section \ref{results}.

\subsection{Discord in the presence of environmental decoherence}
\label{envd}

Environmental decoherence plays a major role in converting a pure quantum state into a statistical mixture. The system, which is strongly coupled to the environment, decoheres during the course of evolution due to interactions with radiation and ambient gas molecules. So, in order to test collapse models, one typically has to go to extreme conditions of very low temperature and pressure such that this effect is negligible. Here, we examine how quantum discord evolves in the presence of environmental decoherence. The master equation has the form 
\begin{equation}
\frac{\partial\rho(x,y,t)}{\partial t}= \frac{i\hbar}{2m}(\nabla_x^2-\nabla_y^2)\rho(x,y,t)-\Gamma(x-y)^2\rho(x,y,t)
\end{equation}
This form was derived in \cite{jooszeh}. Here $\Gamma$ is a parameter that takes into account all the effects due to scattering, emission, absorption of ambient photons and collision with gas molecules. The above form can be compared with Eq. \eqref{masterop} and rewritten as
\begin{equation}
\frac{\partial\hat{\rho}}{\partial t}= -\frac{i}{\hbar}[\hat{H},\hat{\rho}] - \Gamma[\hat{q}^L,[\hat{q}^L,\hat{\rho}]]- \Gamma[\hat{q}^R,[\hat{q}^R,\hat{\rho}]]
\label{envdeco}
\end{equation}
The total contribution to $\Gamma$ comes from scattering ($\Gamma_{sc}$), emission ($\Gamma_{em}$) and absorption ($\Gamma_{abs}$) of photons and collision with gas molecules ($\Gamma_{coll}$) i.e.
$$\Gamma = \Gamma_{sc}+\Gamma_{em}+\Gamma_{abs}+\Gamma_{coll}$$
The effects of scattering, emission and absorption of photons at an ambient temperature $T$ and internal temperature $T_i$ were discussed in \cite{romero-isart} and we quote the results below: 
\begin{eqnarray}
&&\Gamma_{sc} = \frac{8!8 \zeta(9) c R'^6}{9\pi}\left(\frac{k_B T}{\hbar c}\right)^9 \left[{\rm Re}\left(\frac{\epsilon-1}{\epsilon+2}\right)\right]^2 \nonumber \\
&&\Gamma_{em} = \frac{16 \pi^5 c R'^6}{189}\left(\frac{k_B T}{\hbar c}\right)^6 {\rm Im}\left(\frac{\epsilon-1}{\epsilon+2}\right)\nonumber \\
&&\Gamma_{abs} = \frac{16 \pi^5 c R'^6}{189}\left(\frac{k_B T_i}{\hbar c}\right)^6 {\rm Im}\left(\frac{\epsilon-1}{\epsilon+2}\right)
\label{radiation}
\end{eqnarray}
Here $\epsilon$ is the dielectric constant of the material and $k_B$ is Boltzmann's constant. When the system under study is in equilibrium with the environment, we have $T_i = T$ and in that case, the emission and absorption coefficients, $\Gamma_{em}$ and $\Gamma_{abs}$ respectively, are equal. 

The decoherence parameter due to collision with the ambient gas molecules was discussed in \cite{romero-isart} and the form is,
\begin{equation}
\Gamma_{coll} = \frac{8\sqrt{2\pi}\zeta(3)}{3\zeta(3/2)}\sqrt{m_{gas}}n_{gas} \frac{R'^2}{\hbar^2}(k_B T)^{3/2}
\label{coll}
\end{equation}
where $m_{gas}$ is the molecular mass of the ambient gas and $n_{gas}$ is the number density of the gas which is related to the ambient pressure $P$ and temperature $T$ as
$$ n_{gas}= P/k_B T$$
We can express Eq. $\eqref{coll}$ in terms of the external parameters $P$ and $T$ as follows
\begin{equation}
\Gamma_{coll} = \frac{8\sqrt{2\pi}\zeta(3)}{3\zeta(3/2)}\sqrt{m_{gas}}\frac{R'^2 P}{\hbar^2}(k_B T)^{1/2}
\label{coll1}
\end{equation}
Since the ambient medium consists mostly of nitrogen (N$_2$), one can take $m_{gas} = 28.97$ amu.
 
If we compare Eq. $\eqref{envdeco}$ with Eq. $\eqref{master2}$, we see that, in this case
$$\Gamma \equiv 2 \eta$$
As done in earlier sections, the decay rate for the environmental decoherence is now given by,
\begin{equation}
\Lambda_{env} = \frac{\Gamma \hbar}{3 m_0\omega}
\label{lamenv}
\end{equation}
The discord for this model is same as Eq. \eqref{discord} with $\Lambda$ given by $\Lambda_{env}$.

In the present experimental set-up this effect is not important \cite{science,lee} as the environmental noise cannot couple to the phonon modes strong enough due to the very high frequency of phonons ($\sim 40$ THz), and hence will not be considered further in this paper. In the diamond experiment (as pointed out in the reference [22]) the relevant decoherence sources are: phonon-phonon scattering, phonon-electron scattering, phonon scattering due to impurities (mainly Nitrogen impurity) and dislocation. These decoherence effects will have to be reduced, so that entanglement lasts longer than $10^{-12}$ sec. If that can be done, more stringent constraints can be put on the CSL model. However, the results of this section can be used for comparing different models in experiments where environmental decoherence plays an important role and needs to be reduced for detecting the collapse and fundamental decoherence models.\\

\section{Results and Discussion}
\label{results}
In this section, we will analyse (analytically and graphically) the behaviour of discord with time for the present experimental set-up in the presence of CSL collapse mechanism and gravity induced decoherence.

In order to evaluate the discord, we need to consider some standard values for the model and experimental parameters present in Eqs. \eqref{eta}, \eqref{lambda}, \eqref{lamdiosi} and use them in Eq. \eqref{discord}. Taking clue from \cite{science} and \cite{diamondbassi} regarding the values of the experimental parameters $\omega$, $m$, $R$, $d$ and $N$, the corresponding values that we have considered are given in Table \ref{table}. Here, the number density of each diamond crystal is $176.2 \times 10^{27} {\rm m}^{-3}$. Note that, whereas the actual experiment has been done using cylindrical diamond crystals, for obtaining the discord in the presence of gravity induced decoherence, we have considered spherical diamond crystals instead. This has been done in order to simplify the calculations for these models. In order to obtain the radius $R'$ of the spherical diamond crystal, we have kept the volume and the number density of the spherical diamond crystal same as the cylindrical one.

\begin{figure}[H]
\centering
  \includegraphics[width=8cm,height=6cm] {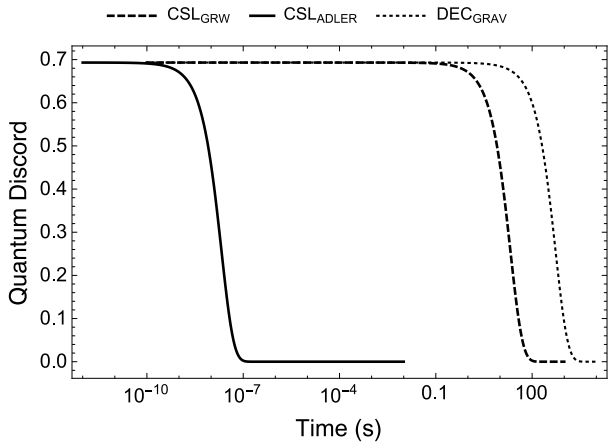}
\caption[Optional caption for list of figures]{\scriptsize Time evolution of quantum discord for CSL (Adler and GRW parameters) and gravity induced decoherence. The smooth (CSL$_{\rm ADLER}$) and dashed (CSL$_{\rm GRW}$) curves represent the quantum discord for CSL with Adler and GRW parameters respectively. The dotted (DEC$_{\rm GRAV}$) curve corresponds to the discord for gravity induced decoherence. The values of the parameters used for plotting the curves are given in Table \ref{table}.} 
\label{final}
\end{figure}

In Fig. \ref{final}, we have plotted the time evolution of quantum discord for all the three models given in Table \ref{table}. From the figure we can infer that in order to detect CSL with Adler collapse parameter, we need to perform the experiment for a time $t > 10^{-7}$ s. Similarly, we can say that CSL with GRW collapse parameter can be detected if we conduct the experiment for $t> 100$ s. For detecting gravity induced decoherence, the experimental time should be atleast $1500$ sec. The actual experiment was performed for approximately $10^{-12}$ s. Hence we need to increase the experimental time significantly in order to detect any of these effects. In particular, detection of CSL with GRW parameter and gravity induced decoherence seems to be very challenging at present as the experimental times required are large.
\begin{widetext}
\begin{center}
\begin{table}[H]
\begin{center}
{\bf {\large Decay rate constants for different models}}
\end{center}
\vspace{0.2in}
\centering
\begin{tabular}{|l|l|l|l|}
\hline
 {\rm {\bf Parameters}}  \quad \quad \quad \quad \quad \quad & ${\rm {\bf CSL_{GRW}}}$ \quad\quad \quad \quad \quad \quad &  ${\rm {\bf CSL_{Adler}}}$  \quad \quad \quad \quad \quad \quad & {\rm {\bf Gravity\ induced}} \\ [1ex]
 \quad\quad\quad \quad \quad \quad & \quad \quad \quad \quad \quad \quad & \quad\quad \quad \quad \quad \quad &  {\rm {\bf decoherence}} \\ [1ex]
\hline
$\omega (s^{-1})$ & $10^{13}$ & $10^{13}$ & $10^{13}$ \\ 
$m(kg)$ &  $10^{-11}$ &  $10^{-11}$ &  $10^{-11}$ \\ 
$N$ &  $5\times 10^{14}$  &  $5\times 10^{14}$ &  $5\times 10^{14}$ \\
$R(\mu m)$ &  $1.3427$ &  $1.3427$ &  $-$ \\ 
$d(mm)$ &  $0.25$ &  $0.25$ &  $-$ \\
$R'(\mu m)$ & $-$ &  $-$ &  $6.97$ \\
$\lambda_{{\rm CSL}}(s^{-1})$ & $10^{-17}$ &  $10^{-8}$ &  $-$ \\
$r_C (m)$ & $10^{-7}$ &  $10^{-7}$ &  $-$ \\
$\Lambda_{{\rm CSL}/Diosi}(s^{-1})$ & $0.0050103$ &  $0.50103\times 10^7$ &  $0.00019659$ \\[1ex]
\hline 
\end{tabular}
\caption{Table showing the different parameter values used to evaluate the decay rate constants for collapse and gravity induced decoherence models. The parameters describing the diamond crystals are taken from the experiment described in \cite{science} and also in \cite{diamondbassi}. The crystals were taken of cylindrical shape originally in the experiment. For simplicity, while calculating the decoherence effect, we have taken it to be a sphere of radius $R'$, keeping the mass and volume of the crystals same as in the original experiment. The symbols used here have already been defined in the text. These parameter values have been used for plotting the graphs and comparing discord for different models.}
\label{table}
\end{table}
\end{center}
\end{widetext}

Note that, if we evaluate the quantum discord for the present experimental setup using pure Schr\"{o}dinger evolution ($\Lambda=0$), we find that the system always remains in a pure state with the quantum discord $\delta = \ln 2$. Thus, simple Schr\"{o}dinger evolution cannot reduce the quantum discord.

\subsection*{Bound on CSL parameters using quantum discord}

\begin{figure}[h!]
\centering
  \includegraphics[width=8cm,height=6cm] {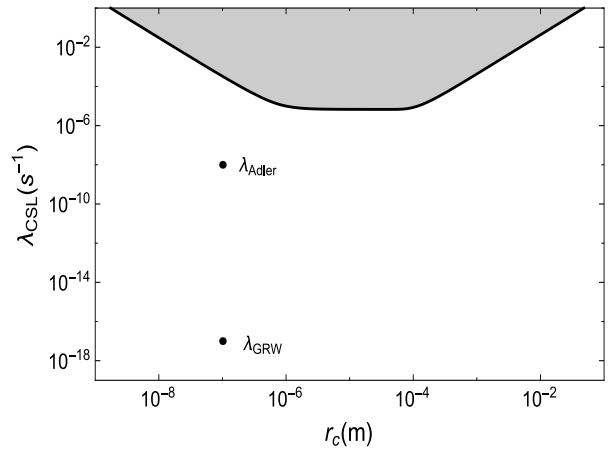}
\caption[Optional caption for list of figures]{\scriptsize Figure shows the allowed $\lambda_{CSL}-r_{C}$ parameter space. The region marked in grey shows the excluded parameter space as obtained from this experiment. For reference, we have also shown the parameter values proposed by Adler ($\lambda_{Adler}$) and GRW ($\lambda_{GRW}$).} 
\label{bound}
\end{figure}

In this experiment, quantum entanglement was observed for $10^{-12}$ sec. Therefore, we expect that the effect of CSL does not set in within this time, which ensures a non-zero quantum discord. Thus, the collapse parameters in CSL should be such that they make the discord zero for a time $t>10^{-12}$ sec which is equivalent to having $\Lambda_{CSL}< 10^{12} s^{-1}$. This gives an upper bound on the CSL parameters. Figure \ref{bound} shows the allowed $\lambda_{CSL}-r_C$ parameter space. This has been obtained by setting $\Lambda_{CSL} = 10^{12} s^{-1}$ and using Eqs. \eqref{lambda}, \eqref{eta}. The values of the constants used are those given in table \ref{table}. The shaded region represents the excluded $\lambda_{CSL}-r_C$ parameter space. Similar bounds were also obtained in \cite{diamondbassi} from the study of quantum entanglement in the same system, thus showing discord is a reliable measure of CSL induced onset of classicality. Also from figure \ref{bound}, we note that for the standard value of $r_C = 10^{-7}$ m, the upper bound on $\lambda_{CSL}\sim 10^{-3}s^{-1}$. 

\vspace{0.1in}

\label{conclu}

\section*{Acknowledgements}
The authors would like to thank Angelo Bassi, Hendrik Ulbricht and Sandro Donadi for helpful discussions.

\end{document}